\input harvmac.tex



\def\unlockat{\catcode`\@=11}
\def\lockat{\catcode`\@=12}

\unlockat

\def\newsec#1{\global\advance\secno by1\message{(\the\secno. #1)}
\global\subsecno=0\global\subsubsecno=0\eqnres@t\noindent
{\bf\the\secno. #1}
\writetoca{{\secsym} {#1}}\par\nobreak\medskip\nobreak}
\global\newcount\subsecno \global\subsecno=0
\def\subsec#1{\global\advance\subsecno
by1\message{(\secsym\the\subsecno. #1)}
\ifnum\lastpenalty>9000\else\bigbreak\fi\global\subsubsecno=0
\noindent{\it\secsym\the\subsecno. #1}
\writetoca{\string\quad {\secsym\the\subsecno.} {#1}}
\par\nobreak\medskip\nobreak}
\global\newcount\subsubsecno \global\subsubsecno=0
\def\subsubsec#1{\global\advance\subsubsecno by1
\message{(\secsym\the\subsecno.\the\subsubsecno. #1)}
\ifnum\lastpenalty>9000\else\bigbreak\fi
\noindent\quad{\secsym\the\subsecno.\the\subsubsecno.}{#1}
\writetoca{\string\qquad{\secsym\the\subsecno.\the\subsubsecno.}{#1}}
\par\nobreak\medskip\nobreak}

\def\subsubseclab#1{\DefWarn#1\xdef
#1{\noexpand\hyperref{}{subsubsection}%
{\secsym\the\subsecno.\the\subsubsecno}%
{\secsym\the\subsecno.\the\subsubsecno}}%
\writedef{#1\leftbracket#1}\wrlabeL{#1=#1}}
\lockat

\def\IL{\relax{\rm I\kern-.18em L}}
\def\IH{\relax{\rm I\kern-.18em H}}
\def\IR{\relax{\rm I\kern-.18em R}}
\def\IC{\relax\hbox{$\inbar\kern-.3em{\rm C}$}}
\def\IT{\relax\hbox{$\inbar\kern-.3em{\rm T}$}}
\def\IZ{\relax\ifmmode\mathchoice
{\hbox{\cmss Z\kern-.4em Z}}{\hbox{\cmss Z\kern-.4em Z}}
{\lower.9pt\hbox{\cmsss Z\kern-.4em Z}}
{\lower1.2pt\hbox{\cmsss Z\kern-.4em Z}}\else{\cmss Z\kern-.4em
Z}\fi}
\def\CM {{\cal M}}
\def\CN {{\cal N}}

\def\CD {{\cal D}}
\def\CF {{\cal F}}

\def\CL {{\cal L}}

\def\CO {{\cal O}}

\def\CH {{\cal H}}

\def\CM {{\cal M}}
\def\CN {{\cal N}}

\def\CO {{\cal O}}

\def\CQ {{\cal Q }}

\font\manual=manfnt \def\dbend{\lower3.5pt\hbox{\manual\char127}}

\def\IZ{\relax\ifmmode\mathchoice
{\hbox{\cmss Z\kern-.4em Z}}{\hbox{\cmss Z\kern-.4em Z}}
{\lower.9pt\hbox{\cmsss Z\kern-.4em Z}}
{\lower1.2pt\hbox{\cmsss Z\kern-.4em Z}}\else{\cmss Z\kern-.4em
Z}\fi}
\def\half {{1\over 2}}

\def\CM {{\cal M}}
\def\CN {{\cal N}}

\def\CO {{\cal O}}

\def\CQ {{\cal Q }}


\def\IZ{\relax\ifmmode\mathchoice
{\hbox{\cmss Z\kern-.4em Z}}{\hbox{\cmss Z\kern-.4em Z}}
{\lower.9pt\hbox{\cmsss Z\kern-.4em Z}}
{\lower1.2pt\hbox{\cmsss Z\kern-.4em Z}}\else{\cmss Z\kern-.4em
Z}\fi}
\def\IB{\relax{\rm I\kern-.18em B}}
\def\IC{{\relax\hbox{$\inbar\kern-.3em{\rm C}$}}}
\def\ID{\relax{\rm I\kern-.18em D}}
\def\IE{\relax{\rm I\kern-.18em E}}
\def\IF{\relax{\rm I\kern-.18em F}}
\def\IG{\relax\hbox{$\inbar\kern-.3em{\rm G}$}}
\def\IGa{\relax\hbox{${\rm I}\kern-.18em\Gamma$}}
\def\IH{\relax{\rm I\kern-.18em H}}
\def\II{\relax{\rm I\kern-.18em I}}
\def\IK{\relax{\rm I\kern-.18em K}}
\def\IP{\relax{\rm I\kern-.18em P}}
\def\IQ{\relax\hbox{$\inbar\kern-.3em{\rm Q}$}}

\def\inbar{\,\vrule height1.5ex width.4pt depth0pt}

\font\cmss=cmss10 \font\cmsss=cmss10 at 7pt
\def\IR{\relax{\rm I\kern-.18em R}}

\def\vol{{\rm vol}}


\def\boxit#1{\vbox{\hrule\hbox{\vrule\kern8pt
\vbox{\hbox{\kern8pt}\hbox{\vbox{#1}}\hbox{\kern8pt}}
\kern8pt\vrule}\hrule}}
\def\mathboxit#1{\vbox{\hrule\hbox{\vrule\kern8pt\vbox{\kern8pt
\hbox{$\displaystyle #1$}\kern8pt}\kern8pt\vrule}\hrule}}


\def\inbar{\,\vrule height1.5ex width.4pt depth0pt}

\font\cmss=cmss10 \font\cmsss=cmss10 at 7pt
\def\IR{\relax{\rm I\kern-.18em R}}

\def\vol{{\rm vol}}

\def\gof{\bar\Gamma^0(4)}


\lref\bateman{A. Erdelyi et. al. , {\it Higher
Transcendental Functions, vol. I,
Bateman manuscript project} (1953)
McGraw-Hill}

\lref\borchaut{R. Borcherds,
``Automorphic forms with singularities on
Grassmannians,'' alg-geom/9609022}

\lref\cohen{H. Cohen, ``Sums involving the
values at negative integers of L-functions of
quadratic characters,''
Math. Ann. {\bf 217}(1975) 271}

\lref\donint{S.K. Donaldson, ``Connections,
Cohomology and the intersection forms of
4-manifolds,'' J. Diff. Geom. {\bf 24 }
(1986)275.}

\lref\eg{G. Ellingsrud and L. G\"ottsche,
``Wall-crossing formulas, Bott residue formula and
the Donaldson invariants of rational surfaces,''
alg-geom/9506019.}

\lref\ez{M. Eichler and D. Zagier, {\it The theory of
Jacobi forms}, Birkh\"auser, 1985}

\lref\finstern{R. Fintushel and R.J. Stern,
``The blowup formula for Donaldson invariants,''
alg-geom/9405002; Ann. Math. {\bf 143} (1996) 529.}

\lref\gottsche{L. G\"ottsche, ``Modular forms and Donaldson
invariants for 4-manifolds with $b_+=1$,'' alg-geom/9506018; J. Am. Math. Soc.
{\bf 9}
(1996) 827.}

\lref\gottzag{L. G\"ottsche and D. Zagier,
``Jacobi forms and the structure of Donaldson
invariants for 4-manifolds with $b_+=1$,''
alg-geom/9612020.}

\lref\mw{G. Moore and E. Witten, ``Integration over
the $u$-plane in Donaldson theory," hep-th/9709193.}

\lref\swi{N. Seiberg and E. Witten,
``Electric-magnetic duality, monopole condensation, and confinement in
${\cal N}=2$ supersymmetric Yang-Mills
Theory,''
hep-th/9407087; Nucl. Phys. {\bf B426} (1994) 19}

\lref\swii{N. Seiberg and E. Witten,
``Monopoles, Duality and Chiral Symmetry Breaking in N=2 Supersymmetric QCD,''
hep-th/9408099}

\lref\vw{C. Vafa and E. Witten,
``A strong coupling test of $S$-duality,''
hep-th/9408074; Nucl. Phys. {\bf B431} (1994) 3.}

\lref\monopole{E. Witten, ``Monopoles and
four-manifolds,''  hep-th/9411102; Math. Res. Letters {\bf 1} (1994) 769.}

\lref\witteni{E. Witten, ``On $S$-duality in abelian
gauge theory,'' hep-th/9505186; Selecta Mathematica {\bf 1} (1995) 383.}

\lref\wittk{E. Witten, ``Supersymmetric Yang-Mills theory
on a four-manifold,''  hep-th/9403193;
J. Math. Phys. {\bf 35} (1994) 5101.}

\lref\zagi{D. Zagier, ``Nombres de classes et formes
modulaires de poids 3/2,'' C.R. Acad. Sc. Paris,
{\bf 281A} (1975)883.}
\lref\zagii{F. Hirzebruch and D. Zagier,
``Intersection numbers of curves on Hilbert modular
surfaces and modular forms of Nebentypus,''
Inv. Math. {\bf 36}(1976)57.}

\lref\lns{A. Losev, N. Nekrasov, and S. Shatashvili, ``Issues in
topological gauge theory," hep-th/9711108; ``Testing Seiberg-Witten solution,"
hep-th/9801061.}

\lref\mmone{M. Mari\~no and G. Moore, ``Integrating over the Coulomb branch in
${\cal N}=2$ gauge theory," hep-th/9712062.}

\lref\mmtwo{M. Mari\~no and G. Moore, ``The Donaldson-Witten function for gauge
groups
of rank larger than one," hep-th/9802185.}

\lref\polishchuk{A. Polishchuk, ``Massey and Fukaya products on elliptic
curves," alg-geom/9803017.}

\lref\fandm{R. Friedman and J.W. Morgan,
``Algebraic surfaces and Seiberg-Witten invariants,''
alg-geom/9502026; J. Alg. Geom. {\bf 6} (1997) 445. ``Obstruction bundles,
semiregularity, and Seiberg-Witten
invariants'', alg-geom/9509007.}

\lref\morganbk{J.W. Morgan, {\it The Seiberg-Witten equations and applications
to the topology of smooth four-manifolds}, Princeton University Press, 1996.}

\lref\DoKro{S.K.~ Donaldson and P.B.~ Kronheimer,
{\it The Geometry of Four-Manifolds},
Clarendon Press, Oxford, 1990.}

\lref\FrMor{R. Friedman and J.W. Morgan,
{\it Smooth Four-Manifolds and Complex Surfaces},
Springer Verlag, 1991.}

\lref\tqft{E. Witten,
``Topological Quantum Field Theory,''
Commun. Math. Phys. {\bf 117} (1988)
353.}

\lref\naka{T. Nakatsu and K. Takasaki, ``Whitham-Toda hierarchy and ${\cal
N}=2$
supersymmetric
Yang-Mills theory," hep-th/9509162; Mod. Phys. Lett. {\bf A11} (1996) 157.}

\lref\gorsky{A. Gorsky, A. Marshakov, A. Mironov and A.Morozov, ``RG equations
from
Whitham hierarchy," hep-th/9802007.}

\lref\dpstrong{E. D'Hoker and D.H. Phong, ``Strong coupling expansions of
$SU(N)$ Seiberg-Witten theory," hep-th/9701055; Phys. Lett. {\bf B397} (1997)
94. }

\lref\BSV{M. Bershadsky, V. Sadov, and
C. Vafa, ``D-Branes and Topological Field
Theories,'' Nucl. Phys. {\bf B463} (1996) 420; hep-th/9511222.}

\lref\bjsv{ M. Bershadsky, A. Johansen, V. Sadov, and
C. Vafa, ``Topological Reduction of 4D SYM to 2D $\sigma$--Models,''
hep-th/9501096; Nucl. Phys. {\bf B448} (1995) 166.}

\lref\doncobord{S. Donaldson,
``Irrationality and the $h$-cobordism conjecture,''
J. Diff. Geom. {\bf 26} (1987) 141.}

\lref\donaldsonii{S. Donaldson, ``Floer homology and
algebraic geometry,'' in {\it Vector bundles in
algebraic geometry}, N.J. Hitchin et. al. eds.
Cambridge 1995.}

\lref\floer{{\it The Floer Memorial Volume}, H. Hofer et. al.
eds., Birk\"auser 1995.}

\lref\munoz{V. Mu\~noz, ``Wall-crossing formulae for algebraic surfaces with
$q>0$," alg-geom/9709002.}

\lref\munozfloer{V. Mu\~noz, ``Ring structure of the Floer cohomology of
$\Sigma \times {\bf S}^1$," dg-ga/9710029.}

\lref\munozqu{V. Mu\~noz, ``Quantum cohomology of the moduli space of stable
bundles over a Riemann surface," alg-geom/9711030.}

\lref\liliu{T.J. Li and A. Liu, ``General wall-crossing formula,'' Math. Res.
Lett. {\bf 2} (1995) 797.}

\lref\morganmrowka{J. Morgan and T. Mrowka,
``A note on Donaldson's polynomial invariants,"
Int. Math. Res. Not. {\bf 10}  (1992) 223.}

\lref\okonek{C. Okonek and A. Teleman, ``Seiberg-Witten invariants for
manifolds with $b_2^+=1$, and the universal wall-crossing formula,"
alg-geom/9603003; Int. J. Math. {\bf 7} (1996) 811.}

\lref\takasaki{Kanehisa Takasaki,
``Integrable Hierarchies and Contact Terms in u-plane Integrals of
Topologically Twisted Supersymmetric Gauge Theories,'' hep-th/9803217
}

\lref\verlinde{E. Verlinde, ``Global aspects of electric-magnetic duality,"
hep-th/9506011;
Nucl. Phys. {\bf B455} (1995) 211. }

\lref \pidtyurin{V. Pidstrigach and A. Tyurin, ``Localization of the Donaldson
invariants along Seiberg-Witten classes,'' dg-ga/9507004.}

\lref\feehan{P.M.N. Feehan and  T.G. Leness, ``$PU(2)$ monopoles and relations
between four-manifold invariants," dg-ga/9709022; ``$PU(2)$ monopoles I:
Regularity,
Uhlenbeck compactness, and transversality," dg-ga/9710032; ``$PU(2)$ monopoles
II:
Highest-level singularities and realtions between four-manifold invariants,"
dg-ga/9712005. }

\lref\weil{A. Weil, {\it Elliptic Functions according to
Eisenstein and Kronecker}, Springer-Verlag, 1976}

\Title{\vbox{\baselineskip12pt
\hbox{YCTP-P7 -98 }
\hbox{hep-th/9804104}
}}
{\vbox{\centerline{Donaldson invariants }
\centerline{ }
\centerline{ for   }
\centerline{ }
\centerline{non-simply connected manifolds }}
}
\centerline{Marcos Mari\~no and Gregory Moore}

\bigskip
{\vbox{\centerline{\sl Department of Physics, Yale University}
\vskip2pt
\centerline{\sl New Haven, CT 06520}}
\centerline{ \it marino@genesis5.physics.yale.edu }
\centerline{ \it moore@castalia.physics.yale.edu }

\bigskip
\bigskip
\noindent
We study Coulomb branch (``u-plane'') integrals
for $\CN=2$ supersymmetric $SU(2),SO(3)$
Yang-Mills theory on 4-manifolds $X$ of $b_1(X)>0,
b_2^+(X)=1$. Using wall-crossing arguments
we derive expressions for the Donaldson invariants
for   manifolds with $b_1(X)>0,
b_2^+(X)>0$. Explicit expressions for
$X=\IC P^1 \times F_g$, where $F_g$ is a Riemann
surface of genus $g$ are obtained using Kronecker's
double series identity. The result might be useful
in future studies of quantum cohomology.

\Date{April 11, 1998}

\newsec{Introduction}

The Donaldson invariants of four-manifolds
have been a source of fascination both in
mathematics and in physics. While there has been
much progress in understanding these invariants,
there is more to learn, particularly in terms of the relation
of the invariants to Floer homology
\floer\ and to Gromov-Witten invariants
\donaldsonii\bjsv.  Understanding the invariants
in these contexts leads to the need to understand the
Donaldson invariants for non-simply connected
four-manifolds $X$. Most investigations of Donaldson
invariants have focussed on the case
$\pi_1(X) = 0$. There exists a mathematical
definition in the non-simply connected case
\morganmrowka\ but comparitively little is
known about this case. This paper derives some
new results on Donaldson invariants for 4-manifolds with
first Betti number $b_1(X)>0$. We do not
consider the effects of torsion in $H_*(X;\IZ)$, nor
the effects of a nonabelian fundamental group.

Recently, using Witten's physical approach to
Donaldson theory \tqft\wittk\monopole\witteni ,
a fairly systematic (physical) procedure has been developed
for deriving various properties of the Donaldson
invariants, including wall-crossing and  blowup
formulae, and the relation to Seiberg-Witten
invariants \mw\lns\mmone\mmtwo\takasaki.
The systematic procedure, which begins with
certain integrals over the Coulomb branch of vacua of
an $\CN=2$ SYM theory can be extended to
higher rank gauge groups and to nonsimply
connected manifolds. In
\mw\lns\ partial results were obtained for
nonsimply connected manifolds. In this paper a
more complete treatment is given for the
rank one groups $SU(2), SO(3)$.

Our main results are:

\item{1.} A wall-crossing formula for the Donaldson
invariants in equations $(3.4)$ and $(3.16)$ below.

\item{2.} An expression for the Donaldson invariants in terms of SW invariants.
For manifolds of simple type this is given in equation $(4.17)$ below. It is a
natural generalization of Witten's formula \monopole, obtained in the simply
connected case.

\item{3.} Explicit expressions for the Donaldson invariants for
$X= \IC P^1 \times F_g$ where $F_g$ is a Riemann surface of genus $g$. We give
answers valid in {\it both } the chambers
$\vol(\IC P^1) \rightarrow 0$ and $\vol( F_g) \rightarrow 0$.

The expressions in equations $(5.18), (5.20)$ below
might prove useful in future studies of the Gromov-Witten
invariants of the moduli space of flat connections on
$F_g$.

\newsec{The $u$-plane integral for $b_1>0$.}

In this section we extend and elaborate on the results of section 10 of \mw. We
consider an arbitrary insertion of observables,
using the proposal for the contact terms in \lns.

Consider an $\CN=2$ $SU(2)$ or $SO(3)$ supersymmetric
Yang-Mills theory
\foot{For simplicity we do not include hypermultiplets.}
on a compact oriented 4-manifold $X$ of $b_1(X)>0$.
As explained in \mw\ the Donaldson-Witten generating
function can be written as $Z_{DW}=Z_u + Z_{SW}$ where $Z_u$
is the Coulomb branch integral and $Z_{SW}$ is the
contribution of the Seiberg-Witten invariants.
The Coulomb branch integral  is only
nonvanishing for $b_2^+(X)=1$, but, by a procedure
explained in \mw\mmtwo\ can be taken as the
starting point for a systematic derivation of
$Z_{DW}$. In the case of $b_1(X)>0$ the
Coulomb integral can be
obtained by a simple
generalization of the arguments in
section three of \mw. First of all, the photon partition
function
includes \witteni\ an integration over $b_1$ zero modes of the gauge field
corresponding
to flat connections. These zero modes span the tangent space to a torus of
dimension
$b_1$, ${{\bf T} ^{b_1}}=H^1(X, \IR)/H^1(X, \IZ)$. The zero modes of the
one-forms $\psi$ live in this tangent space.
As a consequence of having these extra zero modes, the photon partition
function is
\eqn\photonnsc{
({\rm Im \tau})^{{1 \over 2} (b_1-1)} \theta_0 (\tau, {\overline \tau}),}
where $\theta_0 (\tau, {\overline \tau})$ is the Siegel-Narain theta function
introduced
in \witteni\verlinde\mw.

The next ingredient comes from the measure for the $\psi$-fields. The expansion
in zero modes
reads $\psi= \sum_{i=1}^{b_1} c_i \beta_i$, where $\beta_i$, $i=1, \dots, b_1$
is an integral  basis of harmonic one-forms, and we identify
${\cal H}^1(X) \simeq H^1 (X, {\IZ})$. The $c_i$ are Grassmann variables. The
measure for the $\psi$ fields is
then:
\eqn\measure{
\prod_{i=1}^{b_1} { dc_i \over  ({\rm Im \tau})^{1 \over 2}} = ({\rm Im
\tau})^{-{b_1 \over 2}}\prod_{i=1}^{b_1} dc_i.}

We can consider the $c_i$ as a basis of one -forms $\beta_i^\sharp \in H^1
({\bf T }^{b_1}, \IZ)$, dual  to $\beta_i \in  H^1 (X, \IZ)$. In this way we
can identify
\eqn\chern{
\psi = \sum_{i=1}^{b_1} \beta_i \otimes \beta_i^\sharp = c_1 ( {\CL}), }
where $\CL$ is the universal flat line bundle over ${\bf T} ^{b_1} \times X$.

Taking into account \photonnsc\ and \measure, we see that the
Coulomb integral (without any insertion of observables) can be written as \mw:
\eqn\coulbrnch{
\eqalign{
Z_u =
2 \int [da \,d\bar a \,d \eta \,d \chi]
  \int_{Pic(X)}  d\psi
\int dD
&
A^\chi B^\sigma y^{-1/2}  \cr
\exp\Biggl[ {1 \over  8 \pi}
\int ({\rm Im }\tau) D \wedge *D \biggr]
\exp\biggl[ - i \pi
\bar \tau \lambda_+^2 - i \pi \tau \lambda_-^2
&
+   \pi i (\lambda,   w_2(X) ) \biggr]
\cr
\exp\Biggl[
 - {i \sqrt{2}   \over  16 \pi } \int {d \bar \tau \over  d \bar a} \eta
\chi\wedge
(D_+ + 4\pi\lambda_+ )
 + {i \sqrt{2}   \over  2^7 \pi } \int {d \tau \over  da}
(\psi\wedge \psi) \wedge
&
( 4\pi\lambda_-  +
  D_+)    \cr
 + {1 \over  3 \cdot 2^{11} \pi i }  \int    {d^2 \tau \over  da^2} \psi\wedge
\psi
\wedge \psi\wedge \psi \biggr],\cr}
}
where $\int_{Pic(X)}$ denotes a sum over line bundles
and an integration over
${\bf T}^{b_1}$. The integration over $\psi$ is understood as
integration of differential forms
on ${\bf T}^{b_1}$. The orientation of the measure of the finite-dimensional
integral \coulbrnch\ corresponds to
a choice of Donaldson orientation of the moduli space of
instantons \DoKro.

Consider now the generating function for an arbitrary insertion of
zero,
one, two and three observables. We introduce the formal sums of
cycles
\eqn\cycles{
\gamma= \sum_{i=1}^{b_1} \zeta_i \delta_i , \,\,\,\ S= \sum_{i=1}^{b_2}
\lambda_i S_i, \,\,\,\
\Sigma^3= \sum_{i=1}^{b_3} \theta_i \Sigma^3_i, }
where $\delta_i$, $i=1, \dots , b_1$, $S_i$, $i=1, \dots, b_2$ and
$\Sigma^3_i$, $i=1, \dots, b_3=b_1$ are a basis of one, two, and three cycles,
respectively. The basis of one-cycles
$\delta_i$ is dual to $\beta_i \in H^1(X, \IZ)$. The $\lambda_i$ are complex
numbers, and $\zeta_i$, $\theta_i$ are Grassmann variables. The insertion of
observables corresponding to
these cycles is
\eqn\obs{
a_1 \int_{\gamma} Ku +a_2\int_S K^2 u + a_3 \int_{\Sigma^3} K^3 u,
}
where $a_1$, $a_2$ and $a_3$ are constants that should be fixed by comparison
to known mathematical
results. The constant for the two-observable has been already fixed in \mw,
$a_2={i / {\sqrt 2} \pi}$. $K$ is the canonical descent operator in the
normalization of \mw , and we have, explicitly:
\eqn\canonical{
\eqalign{
Ku&= { 1 \over 4 {\sqrt 2}} {du \over da} \psi,\cr
K^2u&= { 1 \over 32}{ d^2 u \over da^2} \psi \wedge \psi - { {\sqrt 2} \over 4}
{du \over da}
(F_+ +  D), \cr
K^3u& = {1 \over {\sqrt 2} 2^7} {d ^3 u \over da^3} \psi \wedge \psi   \wedge
\psi - {3 \over 16}
{ d^2 u \over da^2}\psi \wedge (F_+ + D) - { 3 {\sqrt 2} i \over 8} {du \over
da} (2 d\chi -*d\eta).\cr
}}
In addition, we have to take into account various contact terms associated to
the intersecting cycles \mw\lns. These come from intersections of two, three
and four cycles on the manifold $X$. For the intersection of two cycles we have
\eqn\twoint{
\eqalign{
&S\cap S \in H_0 (X, \IZ), \,\,\,\,\  \Sigma^3\cap \gamma \in H_0(X, \IZ), \cr
&\Sigma^3 \cap S \in H_1(X, \IZ), \,\,\,\,\  \Sigma^3\cap \Sigma^3 \in H_2(X,
\IZ).\cr}}
For intersection of three cycles, we have the possibilities
\eqn\threeint{
S\cap \Sigma^3 \cap \Sigma^3 \in H_0 (X, \IZ), \,\,\,\,\
\Sigma^3\cap\Sigma^3\cap\Sigma^3
\in H_1 (X, \IZ),
}
and for intersection of four cycles we only have the possibility
\eqn\fourint{
\Sigma^3\cap\Sigma^3\cap\Sigma^3\cap\Sigma^3 \in H_0 (X, \IZ).}
The contact term corresponding to $S\cap S$ was obtained in \mw. A proposal for
the structure of the contact terms corresponding to more general intersecting
cycles was made
in \lns.  According to this proposal, the contact term associated to the
intersection of
$p$ cycles is given by the appropriate descendant of the $p$-th derivative of
the Seibeg-Witten prepotential with respect to a deformation
parameter $\tau_0$. Moreover \mmone \mmtwo, this deformation parameter $\tau_0$
can be related to the
dynamically generated scale of the theory $\Lambda$ (in the case of the
asymptotically free
theories) or to the microscopic gauge coupling, in the case of self-dual gauge
theories. For
$SU(2)$ $\CN=2$ supersymmetric Yang-Mills theory, the relation between $\tau_0$
and $\Lambda$ is given by $\Lambda^4={\rm e}^{i \pi \tau_0}$. This is related
to the fact
that $\Lambda$ can be identified with the first slow time of the
Toda-Whitham hierarchy
\gorsky.  Following the proposal of \lns\ the contact terms for the various
intersections in \twoint, \threeint\ and \fourint\ are of the form
\eqn\contact{
\eqalign{
&S^2 T(u) + a_{13} T(u) (\Sigma^3 \cap \gamma) + a_{32} \int_{\Sigma^3 \cap S}
KT(u)+ a_{33} \int_{\Sigma^3\cap \Sigma^3}  K^2 T(u) \cr
&+ a_{332} \CF^{(3)}_{\tau_0} (S\cap \Sigma^3 \cap \Sigma^3) + a_{333}
\int_{\Sigma^3\cap\Sigma^3\cap\Sigma^3} K \CF^{(3)}_{\tau_0}  \cr
&+a_{3333}  \CF^{(4)}_{\tau_0}  (\Sigma^3\cap\Sigma^3\cap\Sigma^3\cap\Sigma^3),
\cr}}
where the $a$'s are constants which will be determined below. In this equation,
$\CF^{(p)}_{\tau_0}$ denotes the $p$-th derivative of the prepotential with
respect to $\tau_0$, and $T(u)= (4/\pi i )  \CF^{(2)}_{\tau_0}$. The contact
term for
$S \cap S$ can be written as \lns\mmone:
\eqn\tu{
T(u) = {1 \over 4} \Bigl[ 2u - a {du \over
da} \Bigr]. }
The constants $a_{ij}$, $a_{ijk}$ and $a_{ijkl}$ will be obtained in terms of
$a_i$, $i=1,2,3$, using single-valuedness of the integrand on the $u$-plane.
Notice
that one expects, on physical grounds, that $a_{ij}$ is proportional to $a_i
a_j$, and so on.

We can already plug the observables \obs\ and the corresponding contact terms
\contact\ into the
generating function and write an explicit expression for the $u$-plane
integral. It is
important, however, to check that the resulting expression has good properties
under
duality transformations, that is,  that the integrand is single-valued in the
$u$-plane.
This is not obvious due to the holomorphic ``functions'' that
appear in \canonical\ and \contact. Following the strategy of \mw, we will
first of all
integrate the auxiliary field $D$. Comparing to the expressions in the
simply-connected case,
we see that the new terms coupling to $D$ appear in the combination
\eqn\dfields{
-{i \over 4 \pi} {du \over da} \Bigl[ (4 \pi \lambda_-+D) \wedge \tilde S
\Bigr]
}
where
\eqn\stilde{
\tilde S=  S-{{\sqrt 2} \over 32}{d \tau \over du} \psi \wedge \psi + { 3 \pi
\over 4 i} a_3 {d ^2 u \over da^2} {d a \over du} \psi \wedge
\Sigma^3 -{\sqrt 2} \pi i a_{33} {dT \over du} \Sigma^3 \wedge \Sigma^3,}
and we interpret $\Sigma^3$ as a harmonic one-form using Poincar\'e duality and
the Hodge theorem. To guarantee that the resulting lattice sum over first Chern
classes is
well-behaved under duality transformations,
the two-form $\tilde S$ should be invariant under duality. To
achieve this, we redefine the $\psi$ field as
\eqn\tildepsi{
\tilde \psi= \psi - { 12 \pi \over {\sqrt 2} i }a_3 {d ^2 u \over da^2} {d a
\over d\tau} \Sigma^3.}
Then, if we choose
\eqn\athreebis{
a_{33}= -9 \pi^2 a_3^2}
$\tilde S$ becomes
\eqn\stildered{
\tilde S= S-  {{\sqrt 2} \over 32}{d \tau \over du} \tilde \psi \wedge \tilde
\psi
+ {9 \pi ^3 {\sqrt 2} i \over 4} a_3^2 \Sigma^3 \wedge \Sigma^3.}
We then see that, if $\tilde \psi$ is a modular form of weight $(1,0)$, then
$\tilde S$ is a modular
form of weight zero. Notice that the redefinition of $\psi$ in \tildepsi\ does
not change the $\psi$ measure.  To obtain the above expression for $\tilde S$,
we have taken into account that
\eqn\dtu{
4 \pi i {dT \over da}-\Bigl({d^2u \over da^2}\Bigr)^2 {da \over d\tau}= \pi i
{du \over da}. }

Once this redefinition has been made, the $u$-plane integral involves a lattice
sum identical to
the one in the simply-connected case but with the substitution  $S\rightarrow
\tilde S$, and additional  holomorphic insertions (coming from the observables
and contact terms)
that, once they are reexpressed in terms of $\tilde \psi$ and $\tilde S$,
should be modular forms
of weight zero. This is in fact the case for appropriate choices of the
constants in \contact.  The
computation is lengthy but straightforward. Duality invariance fixes all the
constants in \contact\ in terms of $a_1$, $a_2$ and $a_3$. The final result is:
\eqn\constantsone{
\eqalign{
&a_{13} = -6 \pi ^2 a_1 a_3, \,\,\,\,\,\ a_{32}=-6 {\sqrt 2} \pi i a_3, \cr
&a_{332}= -72 {\sqrt 2} \pi i a_3^2, \,\,\,\,\,\ a_{333}= 36 \pi ^2 i a_3^3,
\cr
&a_{3333}= -216 \pi^3 i a_3^4.\cr}}
The $u$-plane integral is therefore given by:
\eqn\newcpione{
\eqalign{
Z_u = &\int_{\gof \backslash \CH}
{dx dy \over  y^{1/2}}
\mu(\tau) \int_{Pic(X)} d \psi \exp\biggl[ 2 p u + \tilde S^2 \hat T(u)+
{{\sqrt 2}\over 32} u {d \tau\over du} (\tilde S, \tilde \psi^2) \cr &
+ {a_1 \over 4 {\sqrt 2}} {du \over da} \int_X \tilde \psi \wedge [\gamma]- 3
\pi^2
a_1a_3 u (\Sigma^3 \cap \gamma) -{3 \pi i \over 8} a_3{du \over da} (\tilde S,
\tilde \psi \wedge \Sigma^3) \cr & + {3 {\sqrt 2}\pi ^3 i \over 2} a_3^2 u
(\tilde S,
\Sigma^3 \wedge \Sigma^3) + { 7 \over 3 \cdot  2^{10}} u \Bigl( {d\tau \over
du} \Bigr)^2 \tilde \psi^4 - { {\sqrt 2} \pi i \over 64} a_3 {d \tau \over da}
(\tilde \psi^3 \wedge \Sigma^3)\cr &
+ { 9 \pi^3 i \over 64}a_3^2 u {d \tau \over du} (\tilde \psi^2, \Sigma^3
\wedge \Sigma^3)-{9 {\sqrt 2} \pi ^4 \over 256} a_3^3{du \over da} \int_X
\tilde \psi \wedge
\Sigma^3 \wedge \Sigma^3 \wedge \Sigma^3 \cr
&+ {135 \pi^6 \over 16} a_3^4 u
(\Sigma^3 \cap \Sigma^3\cap \Sigma^3 \cap \Sigma^3) \biggr] \Psi(\tilde S)\cr}
}
where
\eqn\nwcnvpione{
\eqalign{
\mu(\tau) & = - {\sqrt{2} \over  2} {da \over  d \tau} A^\chi B^\sigma\cr
\Psi(\tilde S) & = \exp(2i\pi \lambda_0^2)
\exp\bigl[  - { 1 \over  8 \pi y}({d   u \over  d  a})^2
\tilde S_-^2 \bigr]\cr &
\sum_{\lambda\in H^2+ \half w_2(E) }
\exp\biggl[ - i \pi
\bar\tau (\lambda_+)^2 - i \pi   \tau(\lambda_-  )^2
+ \pi i (\lambda -\lambda_0 ,  w_2(X)) \biggr]  \cr
&
\exp\bigl[- i   { d    u \over  d   a} (\tilde S_-,\lambda_-) \bigr]  \biggl[
(\lambda_+ ,\omega) +{i \over  4 \pi y} {d u\over  d a} (\tilde S_+,\omega )
   \biggr] \cr
\hat T(u)&=T(u)+ {1 \over 8 \pi {\rm Im} \tau} \Bigl({du \over da}\Bigr)^2.
\cr}
}
Notice that, in the above exponential, all the terms are modular forms of
weight zero if
$\tilde \psi$ is a modular form of weight $(1,0)$. There is another check one
can do of the
above functions: one can formally assign an $R$-charge to the cycles in such a
way that the observables have $R$-charge zero, namely: $R(\gamma)=-3$,
$R(S)=-2$, $R(\Sigma^3)=-1$.
Taking into account that $R(\psi)=1$, $R(a)=2$, we see that the definitions of
$\tilde \psi$,
$\tilde S$ are consistent with this $R$-charge assignment and that, moreover,
all the
terms in the exponential of \newcpione\ have zero $R$-charge. This is in fact
an example of
Seiberg's trick since the insertions of \obs\contact\ may be regarded as
operators in some
UV theory ({\it e.g.} in a brane configuration).

The remaining constants $a_1, a_3$ will be fixed below
by comparing to certain topological results \munoz.
In principle, the coefficients in \newcpione\ are fixed by
the proposal of equation $(2.14)$ of the first paper in
\lns. Although we find the proposal of Losev, Nekrasov,
and Shatashvili natural,  because of the many conventions
involved we have not checked that the coefficients
derived in \newcpione\ are consistent with their
equation $(2.14)$.

\newsec{Donaldson wall-crossing}

In this section we want to compute the wall-crossing formulae associated to the
region at
infinity of the $u$-plane.  To compare to mathematical results, it is better
to write the expression in terms of $\psi$, $S$. There are two standard
mathematical facts
for manifolds of $b_2^+=1$ that will be very useful in writing the resulting
wall-crossing formula
\munoz. First, for any $\beta_1$, $\beta_2$, $\beta_3$ and $\beta_4$ in
$H^1(X, \IZ)$, one has $\beta_1 \wedge
\beta_2 \wedge\beta_3 \wedge\beta_4 =0$. Second, the image of the map
\eqn\map{
\wedge: H^1(X, \IZ) \otimes H^1(X, \IZ) \longrightarrow  H^2(X,\IZ)
} is generated by a single rational cohomology class $\Sigma$ (not to be
confused with
the three-cycle $\Sigma^3$). We introduce now the antisymmetric matrix $a_{ij}$
associated to the basis
$\beta_i$ of $H^1(X, \IZ)$, $i=1, \dots, b_1$, as $\beta_i \wedge \beta_j =
a_{ij} \Sigma$.
Finally, we introduce the two-form on ${\bf T}^{b_1}$ as
\eqn\volume{
\Omega= \sum_{i<j} a_{ij} \beta^{\sharp} _i \wedge \beta^{\sharp}_j,
}
which does not depend on the choice of basis. This is a volume element for the
torus, hence
\eqn\voltee{
{\rm vol}({\bf T}^{b_1})= \int_{{\bf T}^{b_1}} {\Omega^{b_1/2} \over
(b_1/2)!}.}
(For the manifolds under discussion
  $b_1$ is even.)
As a simple example, if $X={\IC}P^1 \times F_g$, where $F_g$ is a Riemann
surface of genus $g$, then
$\Sigma=[ {\IC}P^1]$ (the Poincar\'e dual to the two-homology class of
${\IC}P^1$) and
${\rm vol}({\bf T}^{b_1})=1$.

With all these ingredients, we can already write the wall-crossing formulae.
Notice that, because
of the first fact, many terms on the $u$-plane integral \nwcnvpione\ vanish. We
will write the formula first for an insertion of two and four observables. In
this case,
a straightforward generalization of the
arguments in \mw\ gives:
\eqn\wctwofour{
 \eqalign{
WC(\lambda)
& = -   32i     (-1)^{ (\lambda-\lambda_0,   w_2(X))   }
e^{2\pi i\lambda_0^2} (8i)^{-b_1/2}\cr
&\cdot \Biggl[q^{- \lambda^2/2} {da \over d\tau}  \Bigl((u^2-1) {d \tau \over
du} \Bigr)^{1-{b_1 \over 2}}
\Biggl( { ( {2i \over  \pi} {du \over  d \tau} )^2 \over  u^2-1 }
\Biggr)^{\sigma/8}
 \exp\biggl\{ 2p u  +  S^2 T(u)
  -  i  (\lambda , S)/h
\biggr\} \cr
&\,\,\,\,\,\,\,\,\  \int_{{\bf T}^{b_1}} d \psi \exp\biggl\{ {i {\sqrt 2} \over
32} \Bigl( {d^2u \over da^2} {S \over 2\pi} + {d \tau \over da} \lambda, \psi^2
\Bigr) \biggr\}
\Biggr]_{q^0}, \cr}
}
where we use the value obtained in \mw\ for the universal constants $\alpha$,
$\beta$, and
\eqn\htau{
h(\tau)={d a \over du} = {1 \over 2} \vartheta_2 \vartheta_3.}
We can actually compute the integral over ${\bf T}^{b_1}$ and give an
expression in terms of
modular forms which generalizes the expressions given in \mw\gottsche\gottzag.
Using the explicit expressions given in \mw\ for the Seiberg-Witten solution in
terms of modular forms, we find that
\eqn\modu{
{d^2u \over da^2}= 4f_1(q), \,\,\,\,\ {d \tau \over da} = {16 i \over \pi}
f_2(q),}
where
\eqn\qforms{
\eqalign{
f_1(q)&= {2 E_2+ \vartheta_2^4 + \vartheta_3^4 \over 3 \vartheta_4^8}= 1+ 24
q^{1/2} + \cdots, \cr
f_2(q)&={\vartheta_2 \vartheta_3 \over  2 \vartheta_4^8}= q^{1/8} + 18 q^{5/8}+
\cdots.\cr}
}
Using \chern\ and \volume\ we find
\eqn\psisq{
\psi^2= 2 (\Sigma \otimes \Omega),}
hence we can perform the integral over ${\bf T}^{b_1}$ to write a very explicit
expression for \wctwofour:
\eqn\wctwofourmod{
 \eqalign{
WC(\lambda)
& = - {i \over 2}  (-1)^{ (\lambda-\lambda_0,   w_2(X))   }
e^{2\pi i\lambda_0^2} 2^{-b_1/4} {\rm vol}({\bf T}^{b_1}) \cr
&\cdot \Biggl[q^{- \lambda^2/2} h(\tau)^{b_1-2}  \vartheta_4 ^{\sigma}
\exp\biggl\{ 2p u  +  S^2 T(u)
  -  i  (\lambda , S)/h
\biggr\} \cr
&\,\,\,\,\,\,\,\ \sum_{b=0}^{b_1/2} {1 \over (8i)^b} {b_1/2 \choose b} f_1(q)^b
f_2(q)^{b_1/2-b-1} (S, \Sigma)^b (\lambda,  \Sigma)^{b_1/2-b}
\Biggr]_{q^0}. \cr}
}
Notice that this result confirms the conjecture in p. 18 of \munoz.
As a check of this expression, and also to fix an overall coefficient depending
on
$b_1$, we will compute the wall-crossing for the correlator $p^r S^{d-2r}$ on a
ruled surface,
where $d$ is half the dimension of the moduli space, and compare it to the
expressions
in \munoz. Introduce the integer cohomology class $\zeta= 2 \lambda$. In
\munoz, Mu\~noz computes the wall-crossing for walls satisfying $\zeta^2=p_1$
and $\zeta^2 = p_1 +4$, where
$p_1$ is the Pontryagin number of the instanton bundle. For $\zeta^2=p_1$, it
is easy to see that
only the first coefficients in the expansion in $q$ contribute to \wctwofourmod,
and there is
no contribution from the contact term $T(u)$. To compare the formulae, notice
that
we have to multiply our wall-crossing expression by $r!(d-2r)!$, as we are
considering the
Donaldson-Witten generating function. We finally obtain,
\eqn\ruled{
\eqalign{
\delta_{\zeta, \lambda_0} (p^r S^{d-2r})
& =  {1 \over 2}  (-1)^{ (\lambda-\lambda_0,   w_2(X))   }(-i)^{b_1/2} 2^{-3
b_1/4 -b-d} (-1)^{r+d} p^r  {\rm vol}({\bf T}^{b_1}) \cr
& \sum_{b=0}^{b_1/2} {(b_1/2)!\over (b_1/2-b)!}{d-2r  \choose b}
(S,\zeta)^{d-2r-b} (S, \zeta)^b
(\zeta, \Sigma)^{b_1/2-b}. \cr}}
If we compare with \munoz, we find perfect agreement except for an overall
factor $1/2$ (a standard discrepancy
between topological and quantum field theory normalizations), and a  factor
$(-i)^{b_1/2} 2^{-3 b_1/4}$.
The latter factor   is due
to the normalization of the fermions in the physical theory. In order to make
the identification
in \chern\ and to use the normalization of topologists, we have to correct the
$\psi$ measure
with an overall factor $i^{b_1/2} 2^{9b_1/4}$. As we will see in the next
section, with this
normalization the above formula gives the right answer for the generalized
Seiberg-Witten
wall-crossing formula of \liliu\okonek. The case $\zeta^2=p_1+4$ also agrees
with \munoz.
In this case the computation is more involved, as one has to take into account
the first
two coefficients in the $q$-expansion of the various functions in
\wctwofourmod.

We consider now an arbitrary insertion of observables associated to one, two,
and
three cycles. We have new contact terms as well as new terms in the integration
over the
torus. To write these in a convenient way, notice that we can use Poincar\'e
duality and the isomorphism $H_1(X, \IZ) \simeq H^1({\bf T}^{b_1}, \IZ)$ to
obtain
a one-form $\Sigma^{3\sharp}$ in $H^1({\bf T}^{b_1}, \IZ)$. Define
\eqn\innerdef{
\iota_{\beta^{\sharp}_k} \Omega= \sum_p a_{kp} \beta_p^{\sharp}.} Using \chern,
we find
\eqn\inner{
\psi \wedge \Sigma^3= (\iota_{\Sigma^{3 \sharp}} \Omega) \Sigma,
}
In the same way, using the isomorphism $H_1(X, \IZ) \simeq H^1(X, \IZ)$ given
by $\delta_i
\rightarrow \beta_i$, we can define $\gamma^{\sharp} = \sum_{i=1}^{b_1} \zeta_i
\beta_i^{\sharp}$, where the $\zeta_i$ were defined in \cycles. We then obtain,
using \chern\ again,
\eqn\one{
\int_{\gamma} \psi  = \gamma^{\sharp}.}

The functions appearing in the $u$-plane integral can be written in terms of
Jacobi theta functions and Eisenstein series as follows,
\eqn\moremod{
{dT \over da} = -8 f_3(q), \,\,\,\,\,\,\ {\CF}^{(3)}_{\tau_0}= -{\pi^2 \over 4}
f_4(q),
}
where
\eqn\moreq{
\eqalign{
f_3(q)&= {1 \over 16 \vartheta_2 \vartheta_3} \biggl[{1 \over 9\vartheta_4^8}
(2E_2 + \vartheta_2^4 + \vartheta_3^4)^2 -1 \biggr]= q^{3/8} + 12 q^{7/8} +
\cdots,\cr
f_4(q) &= {1 \over 8 (\vartheta_2 \vartheta_3)^2} \biggl[ {1 \over 2}
(\vartheta_2^4 + \vartheta_3^4)
-E_2 + {1 \over 54\vartheta_4^8} (2E_2 + \vartheta_2^4 +
\vartheta_3^4)^3\biggr] =
q^{1/ 4} + 8 q^{3/4} + \cdots.\cr}}

The wall-crossing formula now reads,
\eqn\wcgen{
\eqalign{
&WC(\lambda)
 = - {i \over 2}  (-1)^{ (\lambda-\lambda_0,   w_2(X))   }
e^{2\pi i\lambda_0^2} 2^{7b_1/4} (-i \pi)^{b_1/2} \cr
&\cdot \Biggl[q^{- \lambda^2/2} h(\tau)^{b_1-2} f_2(q)^{-1}  \vartheta_4 ^{\sigma}
\exp\biggl\{ 2p u  +  (S^2 - 6\pi^2 a_1a_3 (\Sigma^3 \cap \gamma) )T(u)\cr
&\,\,\,\,\,\,\,\,\,\,\,\,\,\,\,\ +18{\sqrt 2}\pi^3 i a_3^2 (\Sigma^3 \cap
\Sigma^3 \cap S) f_4(q)
-  i  (\lambda , S)/h-72{\sqrt 2} \pi^3 a_3^2 f_3(q) (\Sigma^3 \wedge \Sigma^3,
\lambda)
\biggr\} \cr
&\,\,\,\,\,\,\,\ \int_{{\bf T}^{b_1}} \exp \Bigl[ 2(P(q), \Sigma) \Omega +
(R(q), \Sigma) \iota_{\Sigma^{3\sharp}}\Omega + Q(q) \gamma^{\sharp} \Bigr]
\Biggr]_{q^0}, \cr}
}
where
\eqn\funex{
\eqalign{
P(q)&= {i {\sqrt 2} \over 16 \pi } \Bigl( f_1(q) S  + 8i f_2(q) \lambda \Bigr),
\cr
R(q)&= 3 \pi a_3  \Bigl( 4i f_3(q) S - f_1(q) \lambda \Bigr),\cr
Q(q)&= {{\sqrt 2} a_1\over 8} {1 \over h}. \cr}
}
In the case of ruled surfaces and for $\zeta^2=p_1$, we find again agreement
with the
expressions obtained by Mu\~noz. In fact, comparing to the formula in p.13 of
\munoz, we can
find the values of $a_1$, $a_3$:
\eqn\constants{
a_1= \pi^{-1/2} 2^{3/4} {\rm e}^{-{\pi i \over 4}}, \,\,\,\,\,\,\ a_3={\pi
^{-3/2} \over 6} 2^{1/4}
 {\rm e}^{\pi i \over 4}.}

\newsec{The Seiberg-Witten contribution}
We can now follow the strategy in \mw\ and find the form of the Seiberg-Witten
contribution at the monopole and dyon cusps. We focus on the monopole cusp, as
the contribution at the
dyon cusp can be obtained using the ${\IZ}_2$ symmetry on the $u$-plane. In
fact, as the
functions involved in the monopole contribution are universal and they have
been
obtained in \mw\ in the simply-connected case, we will be able to {\it derive}
the general
wall-crossing formula of \liliu\okonek.

A crucial ingredient in the discussion of the Seiberg-Witten contribution for
non-simply
connected manifolds is that we have to consider generalized Seiberg-Witten
invariants,
{\it i.e.}, we have to consider correlation functions of one-observables.
Recall that the
basic observable in Seiberg-Witten theory is the two-form $a_D$ on ${\cal
M}_{\lambda}$. The
first descendant of $a_D$ (in the topological field theory associated to the
Seiberg-Witten
monopole equations) is $\psi$, which is a one -form on $X$ and also a one-form
on ${\cal M}_{\lambda}$. It can be written as:
\eqn\oneform{
\psi = \sum_{i=1}^{b_1} \nu_i \beta_i,}
where $\beta_i \in H^1(X, \IZ)$, $i=1, \dots, b_1$ is the basis of one-forms
considered  before,
and
\eqn\oneobs{
\nu_i = \int_{\delta_i} \psi}
are the one-observables of Seiberg-Witten theory. The generalized
Seiberg-Witten invariant
associated to the one-forms $\beta_1, \dots, \beta_r$ is defined by
intersection theory on
${\cal M}_{\lambda}$:
\eqn\gensw{
SW(\lambda, \beta_1\wedge \cdots \wedge \beta_r)= \int_{{\CM}_{\lambda}} \nu_1
\wedge \cdots
\wedge \nu_r \wedge a_D^{d_\lambda-r \over 2},
}
where $d_{\lambda}= \lambda^2-(2 \chi + 3 \sigma)/4$ is the virtual dimension
of ${\CM}_{\lambda}$. These generalized invariants (and their wall-crossing)
have been considered in \okonek.

The Seiberg-Witten twisted Lagrangian near $u=1$ (with the monopole fields
included) can be written as \mw\mmone
\eqn\monlag{
\eqalign{
&  \{ {\overline \CQ}, W \} +  {i  \over 16 \pi}  {\tilde \tau}_M F \wedge F
+ p(u) {\rm Tr} R\wedge R  \cr
& \,\,\,\,\,\ + \ell(u) {\rm Tr} R\wedge {\tilde R}
- {i {\sqrt 2} \over 2^7 \cdot \pi } { d {\tilde \tau}_M \over da_D} (\psi
\wedge \psi) \wedge F
\cr
& + { i \over 3 \cdot 2^{11} \pi} { d^2 {\tilde \tau}_M \over da_D^2} \psi
\wedge \psi \wedge\psi \wedge\psi, \cr}
}
and, as we see, the terms which are not $\CQ$-exact do not depend on the
metric.
The exponentiation of the terms involving the densities $ {\rm Tr} R\wedge R $,
${\rm Tr} R\wedge {\tilde R}$ gives, after integration on $X$, the
gravitational factors $P(u)^{\sigma/8}$,
$L(u) ^ {\chi /4}$ considered in \mw. The term involving $F \wedge F$ gives a
factor
$C(u)^{\lambda^2/2}$, where $F= 4 \pi \lambda$ and $C(u)= {\rm e}^{-2 \pi i
{\tilde \tau}_M}$.
The terms $C(u)$, $L(u)$ and $P(u)$ are universal (they do not depend on the
manifold $X$) and they were found in \mw\ using matching of wall-crossing in
the simply-connected case.
They can be written as
\eqn\univ{
\eqalign{
L(u)&= \pi i \alpha^4\Bigl( (u^2-1) {d \tau_D \over du} \Bigr), \cr
 P(u)&= -\pi^2 \beta^8 a_D^{-1}(u^2-1),\cr
C(u)&= {a_D \over q_D},\cr}}
where $q_D={\rm e}^{2 \pi i \tau_D}$. The last relation tells us that the gauge
coupling $\tilde \tau_M$ appearing in \monlag\ is given by
\eqn\smooth{
\tilde \tau_M= \tau_D -{1 \over 2\pi i}  {\rm log} \,\ a_D,
}
and therefore it is smooth at the monopole cusp. This defines the prepotential
${\widetilde {\cal F}}_M(a_D)$ through the equation ${\widetilde {\cal
F}}''_M(a_D)=\tilde \tau_M$.

First we analyze the Seiberg-Witten contribution when only two and four
observables are
introduced. It can be written as
\eqn\swnsc{
\eqalign{
\langle e^{p\CO+I_2(S)}\rangle_{\lambda,u=1}=&
\int_{{\cal M}_\lambda}
2e^{2i\pi(\lambda_0\cdot\lambda+\lambda_0^2)}
C(u)^{\lambda^2/2}P(u)^{\sigma/8} L(u)^{\chi/4} \cr
&\cdot \exp\left(
2pu + i  {du\over da_D}(S, \lambda)
+S^2 T_M(u)\right) \exp \Bigl[(P_M(u), \Sigma)\sum_{i,j=1}^{b_1} a_{ij} \nu_i
\nu_j \Bigr],
}}
where
\eqn\psmooth{
P_M(u)={i {\sqrt 2} \over 32} \Bigl( {d^2u \over da_D^2} {S \over 2\pi} + {d
\tilde \tau_M \over da_D} \lambda\Bigr),
} which is the magnetic version of the function in the ${\bf T}^{b_1}$ integral
of \wctwofour, but
with the smooth coupling constant \smooth. When we expand the exponential
involving the
one-observables, we obtain the generalized Seiberg-Witten invariants with $2b$
insertions, $b=0, \dots, b_1$. The final expression is:
\eqn\swexpand{
\eqalign{
&\langle e^{p\CO+I_2(S)}\rangle_{\lambda,u=1}= \sum_{b=0}^{b_1/2} {1 \over b!}
{\rm Res}_{a_D=0}
\biggr[ 2e^{2i\pi(\lambda_0\cdot\lambda+\lambda_0^2)}
C(u)^{\lambda^2/2}P(u)^{\sigma/8} L(u)^{\chi/4} \cr
& \,\,\,\,\,\ \cdot \exp\left(
2pu + i  {du\over da_D}(S, \lambda)
+S^2 T_M(u)\right)a_D^{-d_{\lambda}/2 +b-1} (P_M(u), \Sigma)^b \biggl] \cr
& \,\,\,\,\,\ \cdot \sum_{i_p, j_p=1}^{b_1} a_{i_1 j_1} \cdots a_{i_b j_b}
SW(\lambda, \beta_{i_1} \wedge \beta_{j_1} \wedge \cdots \wedge \beta_{i_b}
\wedge \beta_{j_b}).
\cr}
}
On the other hand, the wall-crossing formula for the $u$-plane integral near
$u=1$ can be written
as
\eqn\swwc{
\eqalign{
WC(\lambda)=& 2 \pi i 2^{11 b_1 / 4} i^{b_1/2} \alpha^{\chi} \beta^{\sigma}
e^{2i\pi(\lambda_0\cdot\lambda+\lambda_0^2)} {\rm vol} ({\bf T}^{b_1}) \cr
& \cdot {\rm Res}_{a_D=0}  \biggl[ q_D^{-\lambda^2/2}
\Bigl( (u^2-1) { d\tau \over du} \Bigr) ^{\chi/4} (u^2-1)^{\sigma/8}  \cr
&\cdot \,\,\,\,\,\ \exp\left(
2pu + i  {du\over da_D}(S, \lambda)
+S^2 T_M(u)\right)  (P(q_D), \Sigma) ^{b_1/2} \biggr],\cr}}
where we have included the extra factors depending on $b_1$ that we obtained in
the previous section by comparing to the expressions in \munoz. Notice that,
from \smooth, one has
\eqn\rel{
P(q_D)= P_M(u) + {{\sqrt 2}\over 64 \pi} {1 \over a_D} \lambda,
}
and \swwc\ can then be expanded as:
\eqn\swwcexpand{
\eqalign{
&WC(\lambda)= 2 \pi i  \Bigl({i\over \pi}\Bigr) ^{b_1/2} \alpha^{\chi}
\beta^{\sigma} e^{2i\pi(\lambda_0\cdot\lambda+\lambda_0^2)} {\rm vol} ({\bf
T}^{b_1})\cr
&  \,\,\,\,\,\,\ \cdot \sum_{b=0}^{b_1/2}
{b_1/2 \choose b} {\rm Res}_{a_D=0}  \biggl[ q_D^{-\lambda^2/2}
\Bigl( (u^2-1) { d\tau \over du} \Bigr) ^{\chi/4} (u^2-1)^{\sigma/8} \cr
&\,\,\,\,\,\,\,\,\,\,\,\,\ \cdot \exp\left(
2pu + i  {du\over da_D}(S, \lambda)
+S^2 T_M(u)\right)\cr
& \,\,\,\,\,\,\,\,\,\,\,\,\  \cdot 2^{11 b /2}\pi^b (P_M(u), \Sigma) ^b
(\lambda, \Sigma)^{b_1/2-b} a_D^{b-b_1/2} \biggr].\cr}}
 Now we can compare the expressions for wall-crossing obtained from \swexpand\
and
\swwcexpand. Notice again that, to identify the fields $\psi$ with the
one-observables in  \oneobs\ we have to be careful with possible  normalization
factors needed to agree with
the normalization used by topologists. But for the terms with no insertions,
{\it i.e.} $b=0$
in \swexpand, we should be able to obtain the wall-crossing formula for this
Seiberg-Witten
invariant. In fact, we find:
\eqn\wcformula{
WC(SW(\lambda))=(-1)^{b_1/2} (\lambda, \Sigma)^{b_1/2}  {\rm vol} ({\bf
T}^{b_1}).
}
As $\lambda \in H^2(X, \IZ)+ {1 \over 2} w_2(X)$, and $2 \lambda$ is the
determinant line bundle of the corresponding ${\rm Spin}^c$ structure, we find
perfect agreement with \liliu\okonek\ (notice that the wall-crossing formula in
Theorem 1.2 of \liliu\ should have
an extra factor of $2^{-b_1/2}$). For the wall-crossing of Seiberg-Witten
invariants with one-observable
insertions, the general formula of \okonek\ is (in our notation):
\eqn\general{
WC(SW(\lambda, \beta_{1} \wedge  \cdots \wedge \beta_{r} ))= { (-1)^{b_1-r\over
2} \over ({b_1-r \over 2})!} (\lambda, \Sigma)^{b_1-r \over 2}
\int_{{\bf T}^{b_1}} \beta^{\sharp}_{1} \wedge  \cdots \wedge
\beta^{\sharp}_{r}\wedge \Omega^{b_1-r\over 2},}
and we see that, if we introduce a normalization factor $2^{-9/4}\pi ^{-1/2}i$
for the fields
$\psi$, we obtain from the matching of wall-crossing the
expression
\eqn\genwc{
\eqalign{
&\sum_{i_p, j_p=1}^{b} a_{i_1 j_1} \cdots a_{i_b j_b} WC(SW(\lambda,
\beta_{i_1} \wedge \beta_{j_1} \wedge \cdots \wedge \beta_{i_b} \wedge
\beta_{j_b}))=\cr
&\,\,\,\,\,\,\,\,\,\,\,\ 2^b{ (-1)^{b_1/2-b} \over (b_1/2-b)!} (\lambda,
\Sigma)^{b_1/2-b}  {\rm vol} ({\bf T}^{b_1}),\cr}}
again in agreement with \general. The expression \general\ can be actually
derived
by considering general insertions of one and three-observables.

We then see that, in general, the Donaldson invariants for a non-simply
connected manifold should be written in terms of the generalized Seiberg-Witten
invariants, and only in this case
we have consistent formulae for matching of wall-crossing.  When we include
arbitrary observables associated to one- and three-cycles, the above
expressions are more complicated
and we have to take into account the terms involving $\gamma$ and $\Sigma^3$,
as well as the
new contact terms.

For manifolds of $b_2^+>1$, all the contact terms appear in the Seiberg-Witten
contribution. They are given by the expression in \contact, where the
prepotential is now the prepotential
${\widetilde {\cal F}}_M(a_D)$ (notice that, as all the contact terms are
regular at the monopole cusp, we can take as well the dual prepotential ${\cal
F}_D(a_D)$). In general we have a complicated expression which can be written
explicitly using the previous results. In the
simple type case, $d_{\lambda}=0$, the Seiberg-Witten moduli space consists of
a finite
set of points and counting them with appropriate signs we obtain $SW(\lambda)$.
We only have to compute the different functions at $u=1$ (the first term in the
expansion in $a_D$). Using
also \constants, we can already write an explicit expression for the SW
contribution of $\lambda$ to the Donaldson invariants at $u=1$, with an
arbitrary insertion of observables:
\eqn\swsimple{
\eqalign{
&\langle e^{p\CO+I_2(S)+ I_1(\gamma) +I_3(\Sigma^3)  }\rangle_{\lambda,u=1}=
2^{1 + {7 \chi\over 4} + {11\sigma \over 4}}
e^{2i\pi(\lambda_0\cdot\lambda+\lambda_0^2)} \cr
&  \,\,\,\,\,\,\,\,\,\  \cdot \exp \Bigl\{ 2p + {1\over 2} S^2 - (\Sigma^3 \cap
\gamma) -{1 \over 4} (\Sigma^3\wedge\Sigma^3, S) +{1 \over 96} (\Sigma^3 \cap
\Sigma^3 \cap
\Sigma^3 \cap \Sigma^3) \Bigr\} \cr
& \,\,\,\,\,\,\,\,\,\ \cdot \exp \Bigl\{2(S, \lambda) - {1 \over 4} (\Sigma^3
\wedge \Sigma^3, \lambda)\Bigr\} SW(\lambda). \cr}}
We see that the contact terms give, on the one hand, new contributions which
depend
on the intersection theory on $X$. On the other hand, there is a term coming
from the intersection of
the two-cycle $\Sigma^3 \cap \Sigma^3$ with the Poincar\'e dual of the basic
class $\lambda$, as
was suspected in \wittk\  using the cosmic string
approach. Notice that all the coefficients  in \swsimple\ are real, as needed
for consistency.

We can now write an expression for the Donaldson invariants of non-simply
connected manifolds of simple  type, summing over the basic classes and
considering both the monopole and the dyon contributions. Using the ${\bf Z}_2$
symmetry on the $u$-plane, and taking
into account the $R$-charges of the different terms in \swsimple, we obtain:
\foot{ Curiously, the coefficient of $(\Sigma^3)^4$ is
$1/96 = 2^{-5} \cdot 3^{-1}$. The factor of $3$ would seem to
imply divisibility properties by $3$ for certain intersection numbers. These
are easily confirmed in simple examples, but
we did not find a general proof.}
\eqn\donsimple{
\eqalign{
&\langle e^{p\CO+I_2(S)+ I_1(\gamma) +I_3(\Sigma^3)  }\rangle \cr
&=2^{1 + {7 \chi\over 4} + {11\sigma \over 4}} \biggl( \exp \Bigl\{ 2p +
{1\over 2} S^2 - (\Sigma^3 \cap \gamma) -{1 \over 4} (\Sigma^3\wedge\Sigma^3,
S) +{1 \over 96} (\Sigma^3 \cap \Sigma^3 \cap
\Sigma^3 \cap \Sigma^3) \Bigr\} \cr
&\cdot
\sum_{\lambda}e^{2i\pi(\lambda_0\cdot\lambda+\lambda_0^2)}SW(\lambda)\exp
\Bigl\{2(S, \lambda) - {1 \over 4} (\Sigma^3 \wedge \Sigma^3, \lambda)\Bigr\}
\cr
&+ i^{{\chi + \sigma \over 4} - w_2^2(E)} \exp \Bigl\{ -2p - {1\over 2} S^2 +
(\Sigma^3 \cap \gamma) +{1 \over 4} (\Sigma^3\wedge\Sigma^3, S) -{1 \over 96}
(\Sigma^3 \cap \Sigma^3 \cap
\Sigma^3 \cap \Sigma^3) \Bigr\}\cr
& \cdot
\sum_{\lambda}e^{2i\pi(\lambda_0\cdot\lambda+\lambda_0^2)}SW(\lambda)\exp
\Bigl\{-2i(S, \lambda) + {i \over 4} (\Sigma^3 \wedge \Sigma^3,
\lambda)\Bigr\}\biggr). \cr} }

\newsec{Donaldson invariants for product ruled surfaces}

In this section we will use the above results and the
 formulae in
section 8 of \mw\ to
write   general expressions for the Donaldson invariants of product ruled
surfaces
$X= {\IC}P^1 \times F_g$, where $F_g$ is a Riemann surface of genus $g$.
\foot{This case has also been discussed in eq. $(2.15)$ of
\lns.}   For these manifolds, $b_1=2g$, $b_2=2$, $b_2^+=1$, so $\sigma=0$ and
$\chi=4-2b_1$. $H^2(X, \IZ)$ is generated by the cohomology classes
$[{\IC}P^1]$, $[F_g]$, with intersection form $II^{1,1}$. Consider a
general period
point $\omega$ given by
\eqn\period{
\omega (\theta) ={1 \over {\sqrt 2}}({\rm e}^{\theta}[{\IC}P^1]+ {\rm
e}^{-\theta}[F_g] ). }
The standard constant curvature metrics
for $\IC P^1, F_g$ give natural representatives
for $[\IC P^1], [F_g]$. Thus, choosing coordinates
$z\in \IC$ for $\IC P^1$ and representing
$F_g$ (for $g>1$) as a quotient by a Fuchsian group
of the the Poincar\'e disk:
$\CD=\{ w:  \vert w \vert < 1 \} $, we have
\eqn\stanmetrs{
\eqalign{
[\IC P^1] & ={i \over 2\pi (g-1)}   {dw \wedge d\overline w  \over  (1- \vert w
\vert^2 )^2}, \cr
[F_g]& = {i \over 2\pi} { dz \wedge d\overline z  \over  (1+ \vert z \vert^2)^2
}, \cr}
}
A metric with period point
\period\ then has scalar curvature
$8 \pi ( e^{  \theta} -   e^{ - \theta}(g-1)) $ and will hence
be positive for $e^{2 \theta}>g-1$.
By \monopole\ there are no SW contributions
to the Donaldson invariants for this metric. Thus, to compute
the invariants in the chamber corresponding to the limit of a small volume for
${\IC}P^1$ (with
$\theta$ positive and large), we need only evaluate the $u$-plane integral.

The value of the integral depends on the non-abelian magnetic flux $w_2(E)$.
The vanishing argument of section 5 of
$\mw$ applies when
$w_2(E)\cdot [\IC P^1] \not = 0 $, (i.e. there is nontrivial
flux through the small rational fiber). In this case
we have simply that $Z_u = 0$. Now consider $w_2(E)=0$.
Here we can use the computation of section 8 of \mw. To do this, we choose the
primitive null vector $z=[{\IC}P^1]=\Sigma$. The expression derived in \mw\
will be valid as long as $z_+^2=(z, \omega)^2$
is small, which is the case in the chamber under consideration (corresponding
to $\theta$ large and positive). If we consider the expression \newcpione\ on a product ruled surface, we see
that it involves a lattice sum $\Psi (\widetilde S)$ and a term involving the
observables and the measure. From
this last term we can derive an expression generalizing  the holomorphic
function $f$
appearing in
section 8 of \mw:
\eqn\functionf{
\eqalign{
f&={{\sqrt 2} \over 8 \pi i } (8i)^{1-g} \Bigl( (u^2-1) { d \tau \over du}
\Bigr)^{1-g} {d u \over d \tau} \cr
&\cdot  \exp\biggl[ 2 p u + \Bigl(S^2 - {1 \over 8} (S, \Sigma^3 \wedge
\Sigma^3) \Bigr)  T(u)+
{{\sqrt 2}\over 64}a {d \tau\over da} ( S, \tilde \psi^2) \cr &+{a_1\over 4
{\sqrt 2}} {du \over da}
\int_X\tilde \psi \wedge [\gamma]- u (\Sigma^3
\cap \gamma) -{3 \pi i \over 8} a_3{du \over da} (S, \tilde \psi \wedge
\Sigma^3)\cr &- {1\over 12}  u ( S, \Sigma^3 \wedge \Sigma^3)\biggr], \cr}
}
where we have taken into account the relation \stildered, and $a_3$ is given by
\constants.
We also have that $(\tilde S, z)= (S, \Sigma)$.
Using the computation leading to    equation (8.15) of
\mw, we see
that  the Donaldson invariants of product ruled surfaces in the limit of a
small volume for
${\IC}P^1$ and for $w_2 (E)=0$ are given by
\eqn\dongenruled{
-8 {\sqrt 2} \pi i \biggl[ \Bigl(\int_{{\bf T}^{b_1}} d\tilde \psi \,\ f \Bigr)
h \coth \Bigl( i {(S, [{\IC}P^1]) \over 2h} \Bigr)  \biggr]_{q^0}}
When we consider only insertions of zero and two-observables, the integration
over
the torus is easily done and we obtain the following expression for the
Donaldson invariants
of ${\IC}P^1 \times F_g$, which generalizes the expression obtained in
\gottzag\mw\ for
${\IC}P^1\times
{\IC}P^1$:
\eqn\ruledexp{
-16 i \biggl[ h (u^2-1)  {\rm e} ^{2pu + S^2T(u)} \Bigl( 2h^2 f_1(q) (S,
[{\IC}P^1])
\Bigr) ^g \coth \Bigl( i {(S, [{\IC}P^1]) \over 2h} \Bigr)  \biggr]_{q^0}.}

If $w_2(E)=[{\IC}P^1]$, one can analyze the lattice reduction as in \mw\ including this
nonzero flux, which gives extra phases in the lattice theta function. The effect of these phases is
simply to change the $\coth \Bigl( i {(S, [{\IC}P^1]) /2h}\Bigr)$ in \dongenruled\ by $-i \csc \Bigl(  {(S, [{\IC}P^1]) /2h}\Bigr)$, and we then obtain the generating function for ${\IC}P^1 \times F_g$
and with $w_2(E)=[{\IC}P^1]$:
\eqn\ruledexpf{
-16 \biggl[ h (u^2-1)  {\rm e} ^{2pu + S^2T(u)} \Bigl( 2h^2 f_1(q) (S,
[{\IC}P^1])
\Bigr) ^g \csc \Bigl(  {(S, [{\IC}P^1]) \over 2h} \Bigr)  \biggr]_{q^0},}
where we have included only zero and two-observables.

For applications to Floer theory and Gromov-Witten
theory the most interesting chamber is on the
other side of the K\"ahler cone, namely when
$\theta \rightarrow - \infty$ giving a very small
volume to $F_g$. We can obtain the Donaldson invariants
for this chamber by summing over all the wall-crossing
discontinuities. In general, denoting the lift $2 \lambda_0$
of $w_2(E)$ by
$\lambda_0 = \half \epsilon_1 [\IC P^1] + \half \epsilon_2 [F_g]$
with $\epsilon_{1,2} = 0,1$ we have walls at
$\omega\cdot \lambda_{n,m}=0$ where :
\eqn\walls{
\lambda_{n,m} =
 (n+ \half \epsilon_1) [\IC P^1] + (m+ \half \epsilon_2)  [F_g]
}
for $(n+ \half \epsilon_1)(m+ \half \epsilon_2)< 0$,
$n,m\in \IZ$.
\foot{We require $\lambda^2<0$ rather than
$\lambda^2 \leq 0$ because the ``walls'' at
$\lambda^2=0$ are on the light-cone, and are
never crossed when moving from one chamber to another.}

The infinite sums over wall-crossings can be
written explicitly in terms of modular forms using
a result of Kronecker \weil:
\foot{Curiously, this identity showed up
in recent studies of elliptic quantum cohomology,
\polishchuk.}
\eqn\kronecker{
\eqalign{
\sum_{m=1}^\infty\sum_{n=1}^\infty q^{mn} \bigl[
e^{2 \pi i (n \theta_1 + m \theta_2)}
-
e^{-2 \pi i (n \theta_1 + m \theta_2)} \bigr]
& = -1 + {1 \over  1- e^{- 2 \pi i \theta_1}  }
+ {1 \over  1- e^{- 2 \pi i \theta_2}  }\cr
& - i \eta^3(\tau) {\vartheta_1( \theta_1 + \theta_2 \vert \tau) \over
\vartheta_1(\theta_1 \vert \tau) \vartheta_1(\theta_2\vert \tau) }.\cr}
}

We will consider in detail the cases $2\lambda_0= [\IC P^1]$, $2\lambda_0= [\IC
P^1]+[F_g]$,
which are relevant for the Floer cohomology of $F_g \times {\bf S}^1$
\donaldsonii\munozfloer\ as well as to
the quantum cohomology of the moduli space of stable bundles over $F_g$
\bjsv\munozqu. We consider only insertions of zero, one, and two observables
({\it i.e.} we put $\Sigma^3=0$). As $w_2(X)=0$ for product ruled surfaces, the
only $\lambda$-dependent
factors in the wall-crossing expression \wcgen\ are
\eqn\lambdadep{
q^{-\lambda^2/2} \exp \bigl[ -{i\over h} (\tilde S, \lambda)\bigr],}
where $\tilde S$ has been defined in \stilde.
Define now the formal variables
\eqn\variables{\eqalign{
2\pi \theta_1=&{1\over h} ([\IC P^1] ,\tilde S)= {1\over h} ([\IC P^1] ,S), \cr
2\pi \theta_2=-&{1\over h} ([F_g] ,\tilde S)= -{1\over h} ([F_g] ,S)+ {{\sqrt
2} \over 16} {d \tau \over da} \Omega.\cr}}
The sum of wall-crossings can then be written, for $2\lambda_0= [\IC P^1]$, as
\eqn\kroneckertwo{
\eqalign{
-&\sum_{m=1}^\infty\sum_{n=1}^\infty q^{m(n-1/2)} \bigl[
e^{2 \pi i ((n-1/2) \theta_1 + m \theta_2)}
-
e^{-2 \pi i ((n-1/2) \theta_1 + m \theta_2)} \bigr] \cr
& = {i \over 2 \sin(\pi \theta_1) }+ i \eta^3(\tau) {\vartheta_4( \theta_1 +
\theta_2 \vert \tau) \over
\vartheta_1(\theta_1 \vert \tau) \vartheta_4(\theta_2\vert \tau) },\cr}
}
and for  $2\lambda_0= [\IC P^1]+[F_g]$ as
\eqn\kroneckerthree{
\eqalign{
-&\sum_{m=1}^\infty\sum_{n=1}^\infty q^{(m-1/2)(n-1/2)} \bigl[
e^{2 \pi i ((n-1/2) \theta_1 + (m-1/2) \theta_2)}
-
e^{-2 \pi i ((n-1/2) \theta_1 + (m-1/2) \theta_2)} \bigr]\cr &=
   i \eta^3(\tau) {\vartheta_1( \theta_1 + \theta_2 \vert \tau) \over
\vartheta_4(\theta_1 \vert \tau) \vartheta_4(\theta_2\vert \tau) }.\cr}
}
These explicit expressions are obtained from \kronecker\ by shifting
$\theta_1$, $\theta_2$ appropriately. The quotients of theta functions can be
written in terms of Weierstrass $\sigma$ functions, as in the blow-up formulae
of  \finstern:
\eqn\sigmas{\eqalign{
{\vartheta_1(\theta| \tau) \over \vartheta'_1(0|\tau)} =&{1 \over  \omega_2}
{\rm e}^{-\eta_2 \omega_2 \theta^2} \sigma(t),\cr
 {\vartheta_4(\theta| \tau) \over \vartheta_4(0|\tau)} =&{\rm e}^{-\eta_2
\omega_2 \theta^2} \sigma_3(t),\cr}
}
where $t=\omega_2 \theta$ and $\omega_2$ corresponds to the $a$-period of the
Seiberg-Witten curve, $\omega_2 = { 8\pi \over {\sqrt 2}} h$. The value of the
Weierstrass
zeta function at $\omega_2$ can be written in terms of $E_2(\tau)$ as
\eqn\etatwo{
\eta_2= {\pi^2 \over 6 \omega_2} E_2(\tau).}
In fact, the terms in \sigmas\ involving $\eta_2$ cancel the $E_2$ factors in
the wall-crossing
formula, and the resulting expressions are modular forms of weight zero (after
integrating on
${\bf T}^{b_1}$).

The coefficients in the expansion of the
$\sigma$ functions only depend on the zero-observable $u$. After writing the
Seiberg-Witten
elliptic curve  in Weierstrass form, one has that
\eqn\gis{
g_2= {1\over 4} \bigl( {u^2 \over 3} -{1 \over 4} \bigr),
\,\,\,\,\,\ g_3= {1 \over 48} \bigl( {2u^3 \over 9}- { u \over 4} \bigr),}
and the root relevant for $\sigma_3(t)$ is $e_3=-u/12$. We then have the
expansions
\eqn\sigmaexp{\eqalign{
\sigma(t)&= t- {g_2 t^5 \over 2^4 \cdot 3 \cdot 5} + \cdots,\cr
\sigma_3(t)&=1+{u\over 24}t^2+ \cdots. \cr}
}
Using \sigmas\ we find
\eqn\sigmatheta{ \eqalign{
\eta^3(\tau) {\vartheta_4( \theta_1 + \theta_2 \vert \tau) \over
\vartheta_1(\theta_1 \vert \tau) \vartheta_4(\theta_2\vert \tau)}=&
-{4 \over {\sqrt 2}}h {\rm e}^{-2 \eta_2 \omega_2 \theta_1 \theta_2} { \sigma_3
(t_1 + t_2)
\over \sigma (t_1) \sigma_3 (t_2)} ,
\cr
\eta^3(\tau) {\vartheta_1( \theta_1 + \theta_2 \vert \tau) \over
\vartheta_4(\theta_1 \vert \tau) \vartheta_4(\theta_2\vert \tau) } =&
-{{\sqrt 2}\over 4}h {\rm e}^{-2 \eta_2 \omega_2 \theta_1 \theta_2} { \sigma
(t_1 + t_2)
\over \sigma_3 (t_1) \sigma_3 (t_2)}.\cr}
}

Using these expressions, we can obtain an explicit answer for the
Donaldson-Witten function of product ruled surfaces in the chamber where the
volume of $F_g$ is small. For $2 \lambda_0=
 [\IC P^1]$, we have to add the expression for the chamber where
the volume of $[\IC P^1]$ is small, and the infinite sum of wall-crossing
terms. When the volume of $[\IC P^1]$ is small, we have to use \dongenruled\ but with the $-i \csc \Bigl(  {(S, [{\IC}P^1]) /2h}\Bigr)$ function as in \ruledexpf. We then obtain
\eqn\dwcp{
\eqalign{
Z_{DW}^{[C P^1]}=& -  2^{(7g+1)/2} (-i\pi)^g \Biggl[ h^{2g-1} f_2(q)^{-1} {\rm e}^{(2 p
+{S^2\over 3}) u } \cr
& \cdot \int_{{\bf T}^{b_1}} \exp \Biggl\{ {i  {\sqrt 2}  \over 24 \pi} M(q)
(S,  [\IC P^1]) \Omega + Q(q) \gamma^{\sharp}
\Biggr\} \cdot  { \sigma_3 (t_1 + t_2)
\over \sigma (t_1) \sigma_3 (t_2)}  \Biggr]_{q^0},
 \cr}}
where
\eqn\mq{
M(q) = {\vartheta_2^4 + \vartheta_3^4 \over \vartheta_4^8} .}
For $2 \lambda_0=[\IC P^1]+ [F_g]$ we obtain:
\eqn\dwcf{\eqalign{
Z_{DW}^{[ CP^1]+ [F_g]}=&  {{\sqrt 2} \over 8} 2^{7g/2} (-i\pi)^g \biggl[
h^{2g-1} f_2(q)^{-1} {\rm e}^{(2 p +{S^2\over 3} )u} \cr
& \cdot \int_{{\bf T}^{b_1}} \exp \Biggl\{
{i  {\sqrt 2}  \over 24 \pi} M(q) (S,  [\IC P^1]) \Omega + Q(q) \gamma^{\sharp}
\Biggr\} \cdot { \sigma (t_1 + t_2)
\over \sigma_3 (t_1) \sigma_3 (t_2)}  \biggr]_{q^0}.
 \cr}}
As a check of \dwcf, one can easily derive, using the expansion of the $\sigma$ functions, the result
\eqn\doncheck{
\CD ^{[ CP^1]+ [F_1]}((I( \IC P^1))^2)=-2,}
where $\CD^{[ CP^1]+ [F_1]}$ denotes the Donaldson invariant in the chamber ${\rm vol}(F_1) \rightarrow 0$. This agrees with the computation in \donaldsonii.

Recently the proof of the Atiyah conjecture
on the relation of symplectic and instanton
Floer homology has been completed \munozfloer\munozqu. It is
possible that the above expressions can be used to give another
proof of this conjecture.  In fact,
\eqn\genqc{
Z_{DW}^{([ CP^1], [ CP^1]+ [F_g])} = Z_{DW}^{[ CP^1]}+ Z_{DW}^{[ CP^1]+ [F_g]}}
is a generating function for the
Gromov-Witten invariants of the moduli space
of flat connections \donaldsonii\bjsv\munozfloer\munozqu. In the simple case
$g=1$, one
finds that
\eqn\gone{
{\cal D}^{([ CP^1], [ CP^1]+ [F_1])}(1)=-1, \,\,\,\,\   {\cal D}^{([ CP^1], [
CP^1]+ [F_1])}({\cal O})=2,}
where ${\cal O}$ is the zero-observable. Taking into account that the generator
of the quantum cohomology of the moduli
space of flat connections on $F_1$ (or of the Floer cohomology of $F_1 \times
{\bf S}^1$)
is given by $\beta = -4 {\cal O}$, we obtain the relation $\beta=8$, which is
the first
quantum correction to the classical cohomology ring in genus one.

\bigskip
\centerline{\bf Acknowledgements}\nobreak
\bigskip

We would like to thank Y. Ruan and E. Witten for
some discussions.
We would also like to thank V. Mu\~noz  and
Tianjun Li for very useful and clarifying correspondence.
This work  is supported by
DOE grant DE-FG02-92ER40704.

\listrefs
\end